\documentclass [12pt,a4paper]{article}
\usepackage{graphicx}
\usepackage{psfrag}
\usepackage{graphics}
\usepackage{bm}
\usepackage{color}
\usepackage{verbatim,color,ulem}

\newcommand{\gsim}{\raise.3ex\hbox{$>$\kern-.75em\lower1ex\hbox{$\sim$}}}
\newcommand{\lsim}{\raise.3ex\hbox{$<$\kern-.75em\lower1ex\hbox{$\sim$}}}

\begin{document}

\begin{center}
{\Large\textbf{A New Baseline Metric for Cold Dark Matter}}
\\[0.8\baselineskip]
{\Large\textbf{I. The Early Universe}}
\\[\baselineskip]
 {\large {C.C.\ Wong}\textsuperscript{1}},
 {\large {R.J.\ Rivers}\textsuperscript{2*}}
 \\[0.5\baselineskip]
 1. Department of Electrical Engineering,
     City University of Hong Kong. H.K.
 \\
 2*. Abdus Salam Centre for Theoretical Physics, Physics Department,
     Imperial College London, SW7 2AZ, U.K.
\\[0.7\baselineskip]
Corresponding author*.  E-mail: r.rivers@imperial.ac.uk;
\\
	Contributing author: chicwong@cityu.edu.hk.
\end{center}





\begin{abstract}
\noindent In modelling galaxy structure formation, neither orthodox Cold Dark Matter ($\Lambda$CDM) approaches nor canonical  Modified Newtonian Dynamics (MOND)\cite{milgrom} can easily accommodate the appearance of massive galaxies at early times \cite{mcgaugh2024}.  As a new baseline we propose a single-metric universe which fits within the MOND framework in that there is no new matter in the stress-momentum tensor, to reset the cosmological time scale at which cold dark  matter may have a role.
\\\\
The metric interpolates smoothly between the Schwarzschild metric of central masses at small scales and the Friedmann-Lema$\hat{i}$tre-Robertson-Walker (FLRW) metric of an expanding universe at large scales without discrete boundaries. Within the framework of interpolations proposed by Baker \cite{baker} it is {\it unique} where the Newtonian free fall velocity plays the role of peculiar velocity in an expanding background. Our metric looks old-fashioned in that it superficially resembles a vacuole model with all the artificialities  that this implies, but whereas traditional vacuole models do their best to hide boundary effects, all the new physics in our model arises from the smooth early-universe transitioning of one regime to the other.
\\\\
This metric inherits from MOND the existence of an acceleration scale $a_0$  below which Newtonian gravity is modified. This is no longer the constant of traditional MOND \cite{milgrom} but depends on the background matter density, most simply as $a_0\sim \frac{1}{2} H^2 r$, where $r$ is the relevant distance scale.  As a result, there are additional strong  forces at high red-shifts sufficient at galactic scales, we believe, to trigger galaxy formation without needing to invoke the familiar targets of $\Lambda$CDM searches.
\\\\
We have largely restricted ourselves here to galaxy and galaxy-cluster distances. We are agnostic as to whether we need further ingredients at larger scales but, if we do, the metric provides a solid baseline.
\end{abstract}

\section{Introduction}
\noindent
Serious mass discrepancy problems exist in cosmology and astrophysics which were seen initially to involve central gravitating masses.
At galactic scales the anticipated Newtonian behaviour for rotation curves (essentially derived from the non-expanding Schwarzschild metric) based  on visible matter fails \cite{mcgaugh2004}-\cite{lelli}. A popular resolution to this discrepancy is to assume the existence of invisible Cold Dark Matter ($\Lambda$CDM) in the presence of a cosmological constant. However,
$\Lambda$CDM is beset by a list of problems at small scales  (see the review by Weinberg et al. \cite{weinberg}) and large scales (Hubble tension \cite{riess}-\cite{valentino}, $\sigma_8$ tension \cite{bohringer}-\cite{einasto}, phantom dark energy in DESI survey \cite{{karim}}-\cite{ozulker}).
\\\\
However, once we have invoked invisible dark matter for the resolution of rotation curves, we find new problems far removed from the original motivation. In particular, the problem that we address in this paper is that recent data on High-z Clusters by the James Webb Space Telescope (JWST) shows more structure in the universe at earlier times than $\Lambda$CDM predicts  (e.g. the existence of the morphologically mature massive galaxies \cite{ferreira}-\cite{ferreira2} with, to date, the highest redshift galaxies JADES-GS-Z14 at $z\sim 14$ \cite{ stefano} MoM-z14 at $z=14.44$ \cite{naidu}).
\\\\
Despite numerous searches there has been no direct experimental evidence for CDM to date, leading to "a growing sense of 'crisis' in the dark-matter particle community"\cite{bertone}.
The main alternative approach to $\Lambda$CDM for handling rotation curves is to construct variants of Newtonian gravity that change long-range behaviour to obviate the need for additional particles i.e. to propose a new force law without adding new ingredients to the Stress Momentum Tensor. The most familiar example of this approach is   Milgrom's Modified Newtonian Dynamics (MOND) in which \cite{milgrom} the Newtonian
gravitational acceleration is modified when it takes values below some phenomenological scale $a_0$, where, again, the expanding background does not play any role.   This has its own problems. In its original formulation MOND is a non-relativistic phenomenological model which is very successful at galactic distances \cite{lelli} but fails at larger distances \cite{nusser} and in high Newtonian acceleration epochs.
The need for a relativistic framework has generated several MOND variants 
but the high precision observations of gravitational waves from GW170817 \cite{boran} has ruled out most of these models apart from those of \cite{moffat,skordis,skordis1}. In fact, the stringent bound on Lorentz invariance from GRB221009A observations \cite{cao} suggests that any model with violation of Lorentz invariance is suspect. Even then, recent observations \cite{banik} shows that MOND as a universal physical law fails at Wide-Binary scales and also at the Oort cloud scale within our solar system \cite{tremaine}.
 \\\\
Our approach is very simple. Asymptotically, for short distances with central masses, GR provides the Schwarzschild metric which passes all tests to date; the Friedmann-Lema$\hat{i}$tre-Robertson-Walker (FLRW) metric from GR provides a good description at distances larger than $300Mpc$. We have found a new way to bridge the gap.
\\\\
Crossing over from a Schwarzschild metric to the FLRW metric has a long history which predates mass discrepancy problems. A well known solution for zero cosmological constant is the Einstein-Straus vacuole model \cite{einstein} where the Schwarzschild solution for a central mass (e.g. galaxy) can be matched smoothly to a spherical surface of the expanding FLRW background. This can be extended to non-zero cosmological constant \cite{stuchlik,balbinot} for a dust-like cosmological fluid.  As with MOND, there are no new ingredients beyond those of the Schwarzshild and FLRW models separately. This strict vacuole approach has its own specific problems \cite{carrera,mcvittie,nolan} but,
more generally, with a localised boundary we are either in one regime or the other and we cannot directly address mass discrepancy problems, let alone early time galaxy formation.
\\\\
Our approach consists of constructing a new metric which smoothly interpolates between these two regimes without the artificiality of a vacuole boundary, taking both central forces and expansion into account. Uniquely prescribed within a class of Lema$\hat{i}$tre-Tolman (LT) metrics \cite{baker}, there will be non-Newtonian gravitational effects at intermediate scales without any new ingredients to the Stress-Energy Tensor.
Our metric, whose modified Newtonian mechanics we  term VMOND,  uses only conventional baryonic matter within the matter density.
\\\\
For reasons of brevity, in this paper we limit our discussion largely to the implications of VMOND on the galactic scale. Although we anticipate these transitional effects to be small in general, we will argue here
that at very high red-shifts they are important and observable. 
Our main result is to indicate how, at large scales and  early time, when $H(z)\gg H_0$, VMOND can have a major effect on large-scale structure formation by increasing the rates at which fluctuations can develop.
As such, they look to be compatible with the James Webb Space Telescope (JWST) data  which, as we have noted, shows that a significant population of mature galaxies  at very high redshifts.
 \\\\
 As way-markers to this end, we show i) that the effects on Solar System cosmology are sufficiently small to avoid the conventional MOND problems in modelling the Oort cloud \cite{tremaine} and ii) to relax $\Lambda$CDM bounds on the turnaround distances for galactic clusters \cite{pavlidou}. These tests frame our model for the low-$z$ regime and give us confidence for higher-$z$ applications. Further, the metric provides a stepping stone for the observed non-Newtonian galaxy rotation curves, although we shall not show that here.
\\\\
Our metric may not provide a complete answer but we consider it as a baseline for further study.  The formalism, in itself, does not preclude a dark matter component. With MOND uniquely encoded in the metric we do not see MOND and CDM as essentially antagonistic.  We shall indicate that our metric does the heavy work in dealing with the major discrepancies with Newtonian physics at galactic scale without needing the familiar targets of $\Lambda$CDM. As such, it gives a level playing-field on which details of  dark matter at CMB scales  \cite{lelli}, such as non-baryonic dark matter, can be performed if necessary.

\section{The Model}
\noindent The motivation for this paper is Baker's derivation \cite{baker} of the Bona-Stela construction \cite{bona}, essentially another vacuole model, in which a spherical patch containing a central mass is inserted in a flat  FLRW background. Baker  shows how this construction can be derived from a general Lema$\hat{i}$tre-Tolman metric that  interpolates between the asymptotic solutions.
[On occasion we shall cite from Baker using his equation numbers.]
\\\\
We follow Baker's \cite{baker} approach in adopting a Lema$\hat{i}$tre-Tolman (LT) metric
\begin{equation}
ds^2 = c^2 d\tau^2 -e^{2\alpha(\varrho, \tau)} d\varrho^2-e^{2\beta(\varrho, \tau)} d\Omega^2,
\label{LT}
\end{equation}
for a spatially isotropic "flat" space to describe the motion of test particles in the vicinity of  a single  point-mass M placed at the origin in an expanding cosmological background, with coordinate time $\tau$ and comoving distance $\varrho$.
Time-orthogonality  requires $e^{\alpha} =  \beta' e^{\beta}$ where $^{\prime}$ denotes differentiation with respect to $\varrho$, making (\ref{LT})
a function of $\beta$ only. Different solutions for $\beta$ represent different metrics.
\\\\
We stress that, rather than postulate a stress-momentum tensor for a particular cosmic environment and solve Einstein's equations (in the presence of a cosmological constant), we adopt the contrary approach of constraining the metric to match the Schwarzchild and  FLRW metrics at spatial extremes. Our approach, as Baker's, is as follows:
 Initially, we assume a {\it single} mass $M$ at the origin $r=0$ in a universe which, at large distance from it, behaves like an FLRW fluid with cosmological constant $\Lambda$. With $ e^{\beta} = r$ we look to interpolate between (in mixed coordinates), the Schwarzschild-Lema$\hat{i}$tre (SL) metric at short distances,
\begin{equation}
ds^2=c^2d\tau^2-\frac{2GM}{c^2r}d\varrho^2-r^2d\Omega^2,
\end{equation}
for which
\begin{equation}
e^{\beta} = r = [(3/2)\sqrt{2GM/c^2}(\varrho - c\tau)]^{2/3}.
\label{rrho}
\end{equation}
and the Friedmann-Lema$\hat{i}$tre (FL) metric for scale factor $a(\tau)$ at large distances,
\begin{equation}
e^{\beta} = r = a(\tau)\varrho.
\label{flho}
\end{equation}
Here index $1$ denotes the radial coordinate and indices $2,3$ angular coordinates. We use the index $0$ for temporal coordinates whereas Baker \cite{baker} uses the index $4$. Overdot denotes differentiation with respect to $\tau$.
\\\\
Initially we adopt what is essentially Baker's approach in making the simplest extension of the Schwarzschild- Lema$\hat{i}$tre  metric commensurate with the Friedmann- Lema$\hat{i}$tre metric
\begin{eqnarray}
r&=&e^{\beta(\varrho,\tau)} = a(\tau)L(\varrho, \tau) =
\nonumber
\\
&=& a(\tau)[(3/2)\sqrt{2GM/c^2}(\varrho - cT(\tau))]^{2/3}
\label{aL}
\end{eqnarray}
where
\begin{equation}
 T(\tau) =\int_{t_0}^{\tau} \frac{d y}{a(y)^{\eta}}.
\label{defr1}
\end{equation}
 and $\eta$ is to be determined. The prefactor $a(\tau)$ is chosen to be consistent with the Friedmann-Lemaitre metric at large distance $r$.
For $a(\tau) = 1$  we recover the Schwarzschild-Lema$\hat{i}$tre  metric.
%
%
\subsection{Geodesic Equations}
To choose $\eta$ compatible with observation we need both the geodesic and Einstein's equations, which do not always sit comfortably with one another.
\subsubsection*{$\eta = 3/2$: a unique choice:}
Beginning with the geodesic equations, the low-velocity acceleration equation for radial motion is given as
\begin{equation}
\frac{\ddot{r}}{r}=\frac{\ddot{a}}{a}-\frac{1}{a^{2\eta-3}}\frac{GM}{r^3}-H (2-\eta)\sqrt{\frac{2GM}{r^3a^{2\eta-3}}}.
\label{eta}
\end{equation}
where we have used $
\dot{T} = a^{-\eta}(\tau);\:\:\: \ddot{T} =- \eta \dot{T}H.
$
This modifies Newtonian dynamics in two ways: firstly to change the Newtonian potential and secondly, to include an additional potential due to the expansion of the universe.
The case of $\eta = 3/2$ was rejected by Baker \cite{baker} on the grounds that isotropy is only guaranteed in Einstein's equations for $\eta =3$, which is incompatible with solar system data and he pursued the model no further. The fact that isotropy is not guaranteed does not mean that it cannot be implemented, either exactly, or approximately so. See the Appendix for details,
Henceforth we assume $\eta = 3/2$ to recover Newton's potential.
In  particle free fall in the Schwarzschild-Lema$\hat{i}$tre metric, $\varrho$ is taken as a constant and $r$  reduces as the coordinate time $T=\tau$ increases. In the FLRW metric, $\varrho$ is also taken to be constant in a comoving frame. From Baker (Eq. 28) it follows that
\begin{equation}
\frac{\dot{r}}{r} =H-\sqrt{\frac{2GM}{r^3}}.
\label{sdotr}
\end{equation}
This equation   describes a free (zero energy) particle at a distance $r$ from a central mass $M$, following both the Hubble expansion at large distance and Newtonian gravity at small distances. In particular, it states that the {\it free-fall speed} of a particle in the presence of a central point mass and an expanding background is described by adding the (negative) particle free fall speed in the S-L metric and the (positive) particle free fall speed in an expanding FLRW universe. This is reminiscent of the peculiar velocity treatment in Newtonian perturbation theory but is contrary to the normal procedure of adding (negative) particle {\it free fall acceleration} under point mass gravity to the (positive) point mass acceleration in an expanding FLRW universe.
\\\\
Equation (\ref{sdotr}) is one of the key equations of this analysis. Since we began this programme recent work by de Haas \cite{dehaas} has shown that the same addition equation follows from a biquartionic algebra that generates a relativistic structure in the language of $Q_g$ rotors. In this theory, the underlying spacetime is Minkowskian. The
free fall velocity $v_r$ due to a gravitational source can be represented by a Lorentz boost with rapidity $\Psi_r(x)$ ($\tanh\Psi_r(x)=\frac{v_r}{c}$). In the case of a central point mass, this Lorentz boost is shown to lead to a metric which, in the slow speed limit, reduces to the Painlev\'e-Gullstrand (PG) form of the Schwarzschild metric. The same can be applied to the free fall velocity due to an expanding background. Given a central mass and an expanding background, the theory assumes that the resulting metric (in slow speed limit) can be obtained by the Lorentz boost with additive free fall velocities of Eq.(\ref{sdotr}).
\\\\
We first check that Eq.(\ref{sdotr}) makes sense for small-$z$.

\subsection{Turnaround:}
A particular consequence is that
the condition $\dot{r} = 0$ defines the {\it turnaround radius $r_{TA}$}
\begin{equation}
r_{TA}^3 = \frac{2GM}{H^2},
\label{TA}
\end{equation}
where the two effects balance.
\\\\
For low-z there are observational constraints on turnaround scales, which we now consider.
\subsubsection*{VMOND v. Canonical MOND}
 We  first contrast our model to canonical MOND  for which the turnaround distance is \cite{milgrom}
\begin{equation}
r_{MTA}^2 = \frac{GM}{a_0},
\label{MTA}
\end{equation}
where $a_0=1.2^{+0.4}_{-0.5} \times 10^{-10} ms^{-2}$.
\\\\
For the Solar system with $M = M_{\odot}$, canonical MOND \cite{milgrom} gives  $r_{MTA}\sim 7000 AU$.
Recent work \cite{tremaine} looks at objects travelling in the Oort cloud at distances from $2000AU$  to $50000AU$ from the sun. At $50000 AU$, canonical MOND should impose a strong non-Newtonian acceleration effect on the orbits. Although there are several qualifications, the observations seems to be in conflict with this prediction.
\\\\
However, in our model (\ref{TA}), the distance which separates Newton dominance from non-Newton dominance is
$2.44 \times 10^7 AU$, two orders of magnitude larger than the solar system scale $\sim 10^5 AU$,
assuming the solar system is formed near $H=H_0$. The entire solar system is Newtonian acceleration dominant, including the Oort clouds, in better agreement with the data \cite{tremaine}.
\\\\
The acceleration $a_0$ is the acceleration below which, in canonical MOND \cite{milgrom}, non-Newtonian behaviour takes place. We stress that although it is too simplistic to replace Milgrom's $a_0$ by $\frac{1}{2}H^2r$ in general, where $r$ is the relevant length-scale, we expect a significant divergence from canonical MOND for high redshift where the Hubble constant is large. In Section 4. we suggest that it could provide the strong gravity at high redshift that the JWST galactic results are alluding to.
\subsubsection*{VMOND v. $\Lambda$CDM}
 We return to the geodesic equations in the presence of a cosmological constant, in which the the VMOND turnaround distance is
 \begin{equation}
r_{TA}^3 = \frac{2GM}{H_m^2+H_{\Lambda}^2},
\label{TAH}
\end{equation}
There are experimental bounds on the the turnaround distances for galactic clusters and large galaxies and it is crucial that our model complies with them. In $\Lambda$CDM  calculations the authors of \cite{pavlidou} present three arguments for a theoretical upper limit for the turnaround distance $r^{\rm max}_{\Lambda CDM}$  for such systems. Independent of matter density and time, it takes the form
 \begin{equation}
(r^{\rm max}_{\Lambda CDM})^3 = \frac{3GM}{c^2\Lambda},
\label{LTA}
\end{equation}
There are difficulties with observation and theoretical difficulties with accommodating non-sphericity of galactic clusters. To some extent, non-sphericity relaxes the bound \cite{pavlidou}. In practice, the bound (\ref{LTA}) is saturated for several galactic clusters, with possible observations that might just violate it.
\\\\
 In the absence of matter ($H_m=0$), our turnaround radius (\ref{TAH}) provides a larger turnaround radius limit
 \begin{equation}
(r^{\rm max}_{TA})^3 = \frac{6GM}{c^2\Lambda} = 2(r^{\rm max}_{\Lambda CDM})^3
\label{TAHM}
\end{equation}
which can accomodate the data easily. However, $H_m\neq 0$.
\\\\
 We note that $r_{TA} = r^{\rm max}_{\Lambda CDM}$. If, as a first guess, we take $r^{\rm max}_{\Lambda CDM}$ as characterising astrophysical data then we require
\begin{equation}
H^2\leq 2H^2_{\Lambda};\:\:\Omega_m(1+z)^3=\Omega_{\Lambda}.
\label{mL}
\end{equation}
 for our turnaround radius $r_{TA}$ to be compatible with the bound.
Taking $\Omega_m=0.05$, then $\Omega_{\Lambda}=0.95$ in a flat universe.
Eq. (\ref{mL}) is automatically satisfied by $z=0$.  In fact, $r_{TA}=r_{\Lambda CDM}^{max}$ occurs at $z= 1.67$, which the Matter (baryon)-Dark Energy equality redshift for the FLRW cosmology with no dark matter.
 \\\\
The cited values for large galactic clusters have sufficiently small $z$ (varying from $z\sim 0.005$ for the Formax Cluster to $z\sim 0.07$ for the Corona Borealis Supercluster)
that Eq.(\ref{mL}) is easily satisfied, i.e. $r_{TA} > r^{\rm max}_{\Lambda CDM}$.

\section{The gravitational potential}
Taking the formalism further, the gravitational potential (the second term of which we call the VMOND potential) now takes the form
\begin{equation}
\Phi(r) = -\frac{GM}{r}+H\sqrt{2GMr}-\frac{1}{2}H^2r^2
\label{phi}
\end{equation}
The acceleration equation (the second term of which we henceforth call the VMOND acceleration) is given by
\begin{equation}
\ddot{r}=-\frac{GM}{r^2} -H\sqrt{\frac{GM}{2r}}+\frac{\ddot{a}}{a}r,\:\:\:\:\:\: \frac{\ddot{a}}{a}=-\frac{1}{2} H_m^2+\frac{c^2\Lambda}{3},
\label{nNewton1}
\end{equation}
where $H_m$ is the Hubble parameter in a baryonic matter only universe.
\\\\
Rewriting Eq.(\ref{nNewton1}) in term of mass density gives
\begin{eqnarray}
\frac{\ddot{r}}{r} &=& -\frac{4\pi G}{3} \rho_M(r) -\frac{4\pi G}{3}\sqrt{\rho_M(r) \rho_H}+\frac{\ddot{a}}{a}r,
\nonumber
\\
&&\frac{\ddot{a}}{a}=-\frac{4\pi G}{3}\rho_m+\frac{c^2\Lambda}{3},
\label{nNewton2}
\end{eqnarray}
where $\rho_M(r)$ is the non-local density $(4\pi/3)\rho_M(r) r^3=M$ and $H^2= (8\pi G/3)\rho_H$.
\\\\
In terms of conformal time $\tau_c $ where $d\tau =a d\tau_c$ and $r = aL \neq a\varrho$, the Lema$\hat{i}$tre-Tolman metric takes the form
\begin{equation}
ds^2=a^2\bigg[(1+2\phi) c^2dt_c^2-\frac{dL^2}{(1+2\phi)}-L^2d\Omega^2\bigg].
\label{cc}
\end{equation}
in curvature coordinates $(t_c, \:L)$ with $\phi= -GM/c^2L$, the Newtonian potential.
When $\phi\ll 1$, this is identical in form to the perturbed FLRW metric in the conformal Newtonian gauge (see Mukhanov \cite{mukhanov})
\begin{equation}
ds^2 =a^2\bigg[ (1+2\phi) c^2dt_c^2-(1-2\phi)dl^2 -l^2d\Omega^2\bigg].
\label{lpt}
\end{equation}
However, there is no requirement for $\phi\ll 1$ in our model (\ref{cc}). The metric (\ref{cc}) is applicable everywhere between the cental mass to the Hubble radius.
We shall pursue these seemingly small differences in metric later in looking at large-$H$ fluctuations, when they turn out to be anything but small.

\section{Einstein's equations:}
To recapitulate, we are given a stress momentum tensor with a point mass and an expanding background with uniform matter density in which the metric around  the point mass is specified by free falling speed $\dot{r}=-\sqrt{2GM/r}$, and the uniform density expanding background metric is specified by $\dot{r}=Hr$.
\\\\
Having constructed an interpolating metric we  identify the mass-energy components. The non-Newtonian acceleration can be written in terms of a density $\rho_{MH}=\sqrt{\rho_M(r)\rho_H}$. However, the original stress momentum tensor only possesses the energy density of a point mass $M$ (vacuum) and the background density $\rho_H$.
The induced non-Newtonian density $\rho_{MH}$ does not come from the underlying stress momentum tensor but is a result of the dynamical condition Eq.(\ref{sdotr}).
We can interpret $\rho_{MH}$ as a pure gravitational effect as discussed above, which has no physical material content. The corresponding pressure will be zero, which means we assign its equation of state to be zero. In a stress momentum only approach, we will have to postulate a collisionless (pressureless) matter density $\rho_{MH}$ in Eq.(\ref{nNewton2}) with its correpsonding potential $\Delta \Phi$.
\\\\
 A uniform background density is an idealisation since it  is largely comprised of galaxies which, locally, are central masses. However, the non-local definition of density  will damp out local fluctuations.
With that caveat, our model displays a consistent way in which $\dot{r}$ can interpolate between the two asymptotic solutions with a single metric, without the need for further matter (or further parameters).
This is the basis for Linear and Newtonian perturbations, which will be a significant part of our results.

%
\subsection{$\Lambda = 0$}
We follow Baker's approach \cite{baker} to Einstein's equations. However, our metric in Eq.(\ref{cc}) is more easily understood as a variant of the perturbed metric in Eq.(\ref{lpt}).
 To understand the role of $\eta=3/2$ better, we first consider Einstein's equations without cosmological constant,
\begin{equation}
G_{\mu\nu}= \kappa T_{\mu\nu},\:\:\:\kappa=\frac{8\pi G}{c^4},
\label{EE}
\end{equation}
where $T_{\mu\nu}$ is the stress-energy tensor of the perfect fluid used to mimic the matter content of the universe. We generalise our point mass M at the origin to the {\it spherically symmetric}  (inertial)  gravitational mass $M(r)$ inside a sphere of radius $r$.
Following Krasi$\acute{n}$ski \cite{krasinski} and Baker \cite{baker}, we work with the rest frame of the observer obtained from the Einstein equation
\begin{equation}
G^{\mu}_{\nu}=\kappa T^{\mu}_{\nu}.
\label{eerf}
\end{equation}
The Einstein equation for $T^0_0$ is
\begin{equation}
8\pi G \rho =8\pi G \frac{T^0_0}{c^2}=\frac{1}{r^2}\frac{\partial}{\partial r} \bigg(2GM(r)\bigg).
\label{t442}
\end{equation}
The Einstein equation for $T^1_1$ (Baker Eq.(14)) is
\begin{equation}
-8\pi G  \frac{P}{c^2}= 8\pi G\frac{T^1_1}{c^2}= \frac{1}{(r^2 \dot{r})}\frac{\partial}{\partial \tau}\bigg(2GM(r)\bigg).
\label{t443}
\end{equation}
\begin{eqnarray}
8\pi GT^2_2 &=& 8\pi G T^3_3 =
 8 \pi G T^1_1+\frac{8\pi G }{2}\bigg(r\frac{\partial T^1_1}{\partial r}\bigg).
\label{ni}
\end{eqnarray}
where $\rho$ is the physical matter density and $-P=T^i_i$ is pressure in the $i$ direction. Here we do not assume isotropy at the outset.
\\\\
To capture the total mass representing the flux of  gravitational field over a 3-sphere, the Einstein equation Eq.(\ref{t442}) for $\Phi$ leads to
\begin{equation}
4\pi G \rho=\nabla \cdot (\nabla \Phi) =4\pi G \bigg(\rho_m + \rho_M +\sqrt{\rho_m \rho_M}\bigg) .
\label{T00}
\end{equation}
where $\rho_M$ is due to the central mass $M$. The non-linearity is expressed more clearly if we write (\ref{T00}) as
\begin{equation}
\nabla \cdot (\nabla \Phi) -4\pi G (\rho_m + \rho_M)) = (4\pi G)\sqrt{\rho_m\rho_M}.
\label{mpoi}
\end{equation}
In Eq.(\ref{T00}), the matter density $\rho_{Mm}(r) =\sqrt{\rho_m\rho_M(r)}$ corresponds to the non-Newtonian acceleration in Eq.(\ref{nNewton1}), which we postulate as a gravitational effect (not made of physical particles and thereby {\it pressureless}).
\\\\
This is to be contrasted with the non-linearity of models like that of the superfluid Dark Matter \cite{khoury}, which take the form
\begin{equation}
\nabla \cdot (F(\Phi)\nabla \Phi) -4\pi G \rho = 0
\end{equation}
where $F(\Phi)$ is induced by additional fields. We emphasise again that, in our model, the stress momentum tensor only possesses a point mass and a matter density whose corresponding pressure terms are zero. Non-linearity is an inevitable consequence of interpolation.
\\\\
Next we consider the effect on the pressure due to time variation of $L$. At large distances, $L$ matches $\varrho$ (after rescaling by a constant) asymptotically, the mass contained inside $L$ remains constant when $a$ expands and there is no pressure. As $\tau$ increases and $L$ shrinks, the cosmological background mass contained in the comoving sphere radius $L/\varrho$ will reduce over time according to Eq.(\ref{defr1}).  The question is: Does this matter outflow (even if it is very small) as $L$ decreases constitute the presence of  pressure?
\\\\
For early time scalar perturbation, where $|\phi |\ll1$ is the comoving gravitational potential, the equivalence of (\ref{cc}) to the FLRW metric in conformal Newtonian gauge \cite{mukhanov}
gives the $G^i_i$ component of the perturbed Einstein equation proportional to
\begin{equation}
\frac{d^2\phi}{dt_c^2} +\frac{6}{t_c} \frac{d\phi}{dt_c}=0,
\end{equation}
in a matter dominated universe where $t_c$ is the conformal time.
Assuming \cite{mukhanov} that the process is adiabatic and pressureless
we obtain the well known solution for shortwave length matter density perturbation  \cite{mukhanov}.
\begin{equation}
\bar{\delta}\equiv\frac{\delta \rho}{\rho}\sim \tau^{2/3},
\end{equation}
[We use $\delta$ to denote differential changes and $\bar\delta$ to denote ratios.]
This is the same solution for the overdensity evolution equation in Newtonian perturbation theory at zero sound speed \cite{mukhanov} \cite{bosch2},
Critically, this shows that the overdensity evolves over time without coupling with the mean cosmological background density $\rho_m$.
So far, the analysis above was for $\Lambda =0$.
\\\\
Baker's insistencs on taking $\eta = 3$ was that isotropy was guaranteed by default. However, for $\eta = 3/2$ isotropy can arise naturally in adiabatic dynamcics. A fuller discussion of this is given in the Appendix to this paper.
\subsection{$\Lambda\neq 0$}
The plausible assumption of adiabatic behaviour is sufficient to give no problems with $\Lambda = 0$. In reality $\Lambda$ is non-zero. Whether there is a problem or not for $\Lambda\neq 0$ will depend even more strongly on the empirical behaviour of cosmological matter.
The Einstein equation Eq.(\ref{EE}), including non-zero cosmological constant, is
\begin{equation}
G_{\mu\nu}-\Lambda g_{\mu \nu} = \frac{8\pi G}{c^4} T_{\mu\nu},
\label{EEL}
\end{equation}
where the metric components $g_{\mu\nu}$ are decided by a given $T_{\mu \nu}$. As long as the non-Newtonian density remains pressureless, the analysis above can be repeated.
\\\\
Our starting point is
\begin{equation}
\frac{\dot{r}}{r}= -\sqrt{\frac{2GM}{r^3}}+H,
\label{H2}
\end{equation}
where $H$ is the Hubble parameter, given by
\begin{equation}
H^2= \frac{8\pi G}{3} \big(\rho_m +\rho_{\Lambda}\big)=\frac{8\pi G}{3}\rho_H = H_m^2+\frac{c^2\Lambda}{3} = H^2_m+H^2_{\Lambda}.
\label{H}
\end{equation}
The Einstein equation for $T^0_0$ is
\begin{equation}
 8\pi G \rho =8\pi G\big(\rho_m+\rho_M + \rho_{MH}\big)=8\pi Gc^2T^0_0-\Lambda c^2,
\end{equation}
where $\rho_{MH}=\sqrt{\rho_M\rho_H }$. Assuming $\rho_{MH}$ is pressureless, the energy density due to the presence of physical matter in the stress momentum tensor is the same. We continue to assume an adiabatic approximation such that a free falling particle will not experience pressure due to its mean matter density background. The (negative) pressure due to the cosmological constant background will continue to exist.
 \\\\
We wish to point out that the problem of transforming a FLRW metric with general background fluid, such as a matter or a photon fluid to Curvature coordinates, remains an open problem \cite{mitra}-\cite{mitra2}. We therefore would not attempt to transform our more involved metric into Curvature coordinates here. The closest we can get to a Curvature coordinates formulation is Eq.(\ref{cc}).
\\\\
As for isotropic and non-adiabatic pressure, in this case, $H=\sqrt{(8 \pi G/3)\rho_m}$ is changed to $H=\sqrt{(8 \pi G/3)\rho_H}$  as described in Eq.(\ref{H}).
If we adopt Eq.(\ref{H2}) and $r=aL$, the earlier analysis carries over here and
 the non-adiabatic pressure is $P/c^2 \sim -\rho_m$ (without the adiabatic approximation)  which remains uniform at all scales given spherical symmetry.  We recall from Eq.(\ref{ni}) that anisotropy depends on $\partial P/\partial r$. With a uniform pressure of the FLRW scenario the predicted observable anisotropy from our simple model vanishes at these scales.
\section{Implications for the creation of large-scale structure}
The agreement with $r_{TA}$ for both the Oort clouds and galactic clusters occurs for small $z$. Although we anticipate the effects of the interpolation to be small in general,  we highlight one potentially important implication of our model; that baryonic overdensities at recombination can evolve fast enough to match observation.
\\\\
One of the major mass discrepancy problems that leads to the dark matter postulate is that the baryonic overdensity  necessary for the creation of large scale structures  due to a Cosmic Microwave Background temperature variation at recombination cannot evolve fast enough to match late time observations.

\subsection{Perturbation theory}
The metric in Eq.(\ref{cc}) indicates an underlying connection of our model with Linear perturbation theory.  As a first step we contrast it to the Newtonian perturbation theory (NPT) that follows from (\ref{lpt}) \cite{mukhanov}-\cite{bosch2}.

\subsubsection*{Newtonian perturbation theory}
In Newtonian perturbation theory (NPT),  the background metric is taken to be the FLRW metric with a mean cosmic background matter density $\bar{\rho}$ where $a(\tau)$ is the scale factor in this expanding background. The physical radius of a free paricle is given in terms of parametrisation of $r$,
\begin{equation}
r=al,\:\:\:
\end{equation}
The Eulerian fixed point frame  velocity
\begin{equation}
u=\dot{a}l+a \dot{l} =Hr+a\dot{l}
\end{equation}
where $\dot{r}=Hr$ is the Hubble's Law in the uniformly expanding background. Note: Here $l$ is not the same as $L$ in Eq.(\ref{aL}).
\\\\
The density perturbation $\delta \rho (r) =\rho(r) -\bar{\rho}$, $\rho(r)$ is the total density at $r$, against the expanding background is represented primarily in terms of peculiar velocity $v_r$ which is related to the momentum density perturbation  due to overdensity $\delta \rho$ through the equation
\begin{equation}
u=Hr+v_r.
\label{2}
\end{equation}
$v_r$ follows its own gravitational potential and is specified up to the Poission equation for $\delta \rho$
\begin{equation}
\nabla_r^2(\Phi_0+ \delta \Phi)=4\pi G (\bar{\rho}+\delta \rho).
\label{NPTpotl}
\end{equation}
where $\Phi_0$ is the potential corresponding to the mean background density $\bar{\rho}$ and $\delta \Phi$ is potential corresponds to the density perturbation $\delta \rho$.
The Euler equation, for negligible entropy and pressure, is given by
\begin{equation}
\frac{D}{D\tau}u=- \bigtriangledown_r (\Phi_0+\delta \Phi),
\label{DD1}
\end{equation}
Going to the Lagrangian frame which follows the individual moving particle using the Lagrangian derivative,
\begin{equation}
\frac{D}{D\tau}=\frac{\partial}{\partial \tau}-u\cdot \nabla_r,\:\:\: \:\:\: \nabla_r=\frac{1}{a}\nabla_l.
\end{equation}
Subtract out the background relation from Eq.(\ref{DD1})
\begin{equation}
\frac{D}{Dt} (Hr)=-\nabla_r \Phi_0,
\end{equation}
and keeping only the first order terms leads to
\begin{equation}
\dot{v}_r+Hv_r=-\frac{1}{a}\nabla_l \delta\Phi.
\label{euler11}
\end{equation}
Combining these two equations into the continuity equation
\begin{equation}
\frac{D}{D\tau} \rho =-\nabla_r \cdot (\bar{\rho}+\delta \rho) u
\end{equation}
where $\rho v_r$ is the "momentum density fluctuation" due to density perturbation $\delta \rho$.
\\\\
Assuming matter dominance $\rho =\bar{\rho}+\delta \rho \propto a^{-3}$, $\dot{\rho}=-3H\rho$ and $\bar{\delta} =\delta\rho/\rho$ and keeping the linear order terms, one obtain
\begin{equation}
\frac{\partial}{\partial \tau} \bar{\delta}-\frac{1}{a} \nabla_l \cdot v_r=0
\label{12}
\end{equation}
Differentiate Eq.(\ref{12}) w.r.t. to $\tau$ and apply ($\nabla_r \cdot$) to the Eq.(\ref{euler11}) and combine with the Poisson equation Eq.(\ref{NPTpotl}),
one obtains the overdensity evolution equation
\begin{equation}
\ddot{\bar{\delta}}+2H\dot{\bar{\delta}} +4\pi G \bar{\delta} \rho=0
\label{5}
\end{equation}
which is the pillar of structure formation evolution. Although the above formalism is well known
in the literature, we recall them so that the implication of the additional acceleration and additional potential in VMOND can be conveniently evaluated.
\\\\
We know in practice, for example in the $\Lambda$CDM model, that the perturbation growth equation Eq.(\ref{5}) based on NPT is usually extrapolated to $\delta\sim 1$ in the direct collapse model \cite{binney}-\cite{bromm}. (In \cite{hwang}, it is also argued using non-linear perturbation that the Newtonian perturbation theory can be extended to the non-linear region.) However, when comparing to data, something is still missing after the inclusion of CDM particles. Replacing baryons by dark particles only changes the initial overdensity value but does not change the following analysis.
\\\\
For example, at recombination, for which $z=1080$, consider a baryon perturbation arising from a source perturbation which forms a comoving shell. For is-entropic perturbation,  we have radiation overdensity $\bar\delta_{rad}= (4/3)\bar\delta_{b}$. Since $\rho_{rad}\propto T^4$, whence  $\bar\delta_{rad}=4\delta T/T$, we have $\bar\delta_b=3\delta T/T$.
 For a comoving shell of order $150Mpc$ the CMB average temperature variation $\delta T/T=1-3\times 10^{-5}$ corresponds to an initial baryon overdensity $\bar\delta_{int} = 3\delta T/T =3-9 \times 10^{-5}$. For $z<1080$, if we had only the Newtonian overdensity, this evolves according to the equation
\begin{equation}
\bar\delta=\bar\delta_{int}\bigg(\frac{  1081}{1+z}\bigg).
\label{delnewton}
\end{equation}
At $z=0$, we have $\bar\delta \sim 3-9\times 10^{-2}$, which is at variance with
the very recent late time observation  $\sqrt{ \langle\bar\delta\rangle^2}=\sigma_8 \sim 0.745$ \cite{bohringer}. [In more detail, this value is the root mean square of the amplitude of
matter perturbations smoothed over $8h^{-1} Mpc$ where $h$ is the
Hubble constant in units of $100 km s^{-1} Mpc^{-1}$. Even the inclusion of an additional biasing factor $b_{gal}=1.3\pm 0.13$ for clustered galaxies \cite{einasto} is not enough to give agreement.] We need something more.
\\\\
In fact, CDM of itself is not enough at galactic scales at very early time.
The James Webb Space Telescope (JWST) recently observed \cite{ferreira}-\cite{yan} large galaxies at very high redshift ($z>11$, that is within $500 Myr$ from recombination) compared to the $2 Gyr$ expected from small-halo merging process of the $\Lambda CDM$ model. A significant number of Supermassive black holes (SMBH) at $10^9M_{\odot}$ are observed at $z\,\,\gsim\,\, 6$ \cite{fan}, which suggests that a large host galaxy should have virialised by $z>10$ (for a black hole seed with mass $10^2-10^5M_{\odot}$) to accrete to the observed mass within the Eddington limit \cite{bromm}.
More recent JWST observations \cite{ferreira2} of around 4000 galaxies show that for large mass galaxies $(\geq 10^{9}M_{\odot})$, the fraction of spiral, spheroid and irregular galaxies are constant over the redshift range $1.5<z<6.5$, which means that these large galaxies are already well developed by $850 Myr$.
\\\\
To account for these observations, an initially expanding overdensity cloud needs to turnaround and decouple from the cosmological background at high redshift. This requires a turnaround redshift much higher than $z\simeq 6$ for the galaxies studied in \cite{ferreira2} and for the SMBH discussed in \cite{bromm}. The dark matter potential at recombination is well known, persisting until $z\sim 4$  \cite{miyatake}. These early large galaxies therefore require a new mechanism to increase the dark matter potential immediately after recombination and subsequently return the dark matter potential to the $\Lambda CDM$ model at $z\sim4$.  
\\\\
Our resolution of the problem is as follows.

\subsubsection*{VMOND perturbation theory}
Starting from the NPT formalism, we take the peculiar velocity perturbation to be the particle free fall velocity $v_r=a \dot{L}$ ($l=L$) corresponding to an effective central mass perturbation  such that
\begin{equation}
u=\dot{r} =Hr+v_r,\:\:\: v_r=-\sqrt{\frac{2GM}{r}}.
\label{11}
\end{equation}
This modified NPT gives us the acceleration equation Eq.(\ref{nNewton1}), leading to the introduction of an  additional MOND-like acceleration and a different overdensity evolution equation valid for $\delta\leq 1$.
\\\\
In a matter dominant expanding universe, for a local baryon overdensity with mass $M = \bar{\delta}\rho_m$, radius $r$,
\begin{equation}
\frac{2GM}{r^3}= \frac{8\pi G}{3}\rho_M=\frac{8\pi G}{3}{\bar\delta} \rho_m.
\end{equation}
 More specifically,
\begin{equation}
M={\bar\delta} \rho_m \frac{4\pi}{3}r^3,\:\;\:H(z)=\sqrt{\frac{8\pi G}{3}\rho_m},\:\:\: \:\frac{\ddot{a}}{a}=-\frac{4\pi G}{3} \rho_m.
\label{h2r3}
\end{equation}
and, including an angular momentum term, Eq.(\ref{nNewton1}) becomes
\begin{equation}
\ddot{r}=\frac{h^2}{r^3} -\big( {\bar\delta}+{\bar\delta}^{1/2} +1\big) \frac{4\pi G}{3}\rho_m(z) r.
\label{ddr}
\end{equation}
The potential $\phi$ in Eq.(\ref{phi}) due to the overdensity now includes an extra (dynamical) matter density ${\bar\delta}^{1/2}\rho_m$. [We note that a "phantom dark matter" density in addition to the Newtonian density is also used in providing MOND gravity by Milgrom \cite{milgrom1986}, \cite{oria}.]
\\\\
This leads to the Einstein equation
\begin{equation}
\nabla^2\Phi  =\nabla^2(\Phi_0+\Delta\Phi+\delta \Phi)=4\pi G \rho=4\pi G (1+\Delta)\rho_m,\:\:\: \Delta ={\bar\delta}+{\bar\delta}^{1/2}.
\label{Dpotl}
\end{equation}
This differs from Eq.(\ref{NPTpotl}) by having an extra potential $\Delta \Phi$ (from Eq.(\ref{phi})) and the corresponding overdensity $\delta^{1/2} \rho_m$.
The Euler equation for the overdensity for negligible spatial entropy gradient takes the form Eq.(\ref{euler11}) and Eq.(\ref{ddr})
\begin{equation}
\dot{v}_r=-4\pi G\delta \rho_m r-4\pi G \sqrt{\delta} \rho_m r,
\end{equation}
where $v_r$ is the radial velocity which is under the influence of both the Newtonian and the non-Newtonian potential.  This radial velocity from Eq.(\ref{sdotr}) will lead to both a Newtonian and non-Newtonian term in its peculiar acceleration Eq.(\ref{nNewton1}) and Eq.(\ref{ddr}).
\\\\
 We also need the continuity equation for $\rho=(1+\delta+\sqrt{\delta}) \rho_m$
\begin{equation}
\frac{\partial \rho}{\partial t}+ \nabla \cdot (\rho v_r)=0,
\label{ceqn}
\end{equation}
which also has a Newtonian perturbation term $\delta\rho_m$ and a non-Newtonian perturbtation term $\delta^{1/2}\rho_m$.
The end result is
\begin{equation}
\ddot{\Delta}+2H\dot{\Delta}-4\pi G\rho_m \Delta=0,
\label{Dmatteronly}
\end{equation}
and the growth mode for $\Delta$ becomes
\begin{equation}
\Delta \propto \tau^{2/3},\:\;\;\,{\rm permitting}\,\,\,\,{\bar\delta} \propto a^2 \propto \tau^{4/3}.
\label{Dmatteronly2}
\end{equation}
for small ${\bar\delta}$, which provides a much faster overdensity $\bar\delta$ evolution after recombination without the need for introducing invisible matter.

\section{Over-density evolution from VMOND}
The question then is whether this initially faster overdensity evolution due to our mechanism is enough to give agreement with data. We shall suggest that it can be, but the argument is not straightforward. We have to make assumptions about galaxy formation, power spectra, simulations  and energy redistribution that take us a long way beyond our original model.
\\\\
From an initial overdensity $\bar\delta_{int}$ at recombination, we can calculate the overdensity $\bar\delta$ at redshift $z<1080$ by
\begin{equation}
\bar\delta+\sqrt{\bar\delta}=\big(\bar\delta_{int}+\sqrt{\bar\delta_{int}}\big) \bigg(\frac{1081} {1+z}\bigg)=\frac{A_0}{1+z},
\label{dpsd}
\end{equation}
Galaxy formation is favoured near the shell's origin and at a radius of $150Mpc$ and at scales that are less than $2Mpc$. In the absence of VMOND the baseline CMB average temperature variation $\delta T/T =1\times 10^{-5}$ corresponds to an initial baryon overdensity $\bar\delta_{int} =3 \times 10^{-5}$ and $A_0=5.94$. In an uniform density approximation, the initial density for a galaxy should be much higher than this $150Mpc$ average  of $10^{-5}$.
\\\\
To account for high redshift large galaxies, one needs an overdensity to turnaround at sufficiently higher redshift. As a point of reference, \cite{bromm} uses a $\bar{\delta}_{int}=10^{-3}$ at $z\sim 1080$ to develop into a galaxy of $10^8 M_{\odot}$. In \cite{sanders}, in a spherical galaxy (with mass $\sim 10^{10.5}M_{\odot}$) formation under MOND, Sanders takes $\bar\delta_{int} =1.8\times 10^{-3}$ which corresponds to $A_0=47.8$.
\\\\
Although Sander's choice is sufficient for most of our purposes, we choose a slightly higher value $\bar{\delta}_{int}=2.8\times 10^{-3}$ ($A_0=60$) since we are interested in Milky Way-like galaxies where we have more data to test our model. In our type of direct collapse model, the more massive a structure, the "lower" the initial overdensity, e.g. a $\sigma_8$ initial overdensity choice close to the CMB overdensity averaged at $\delta\sim 3\times 10^{-5}$.  We obtain the turnaround redshift $z_{ta}$ where $\bar\delta=1$ from Eq.(\ref{dpsd}),
%
\begin{equation}
1+z_{ta}=\frac{A_0}{2}=30, \:\:\: z_{ta}=29.
\label{ta}
\end{equation}
We stress that  this turnaround redshift (resulting from this choice of $\delta_{int}$) is well within an observationally viable  redshift range $15\:\lsim z\: \lsim\: 50$ \cite{barkana}, from 21cm radiation.
\\\\
When comparing to the Newtonian gravity-only evolution Eq.(\ref{delnewton}) gives
\begin{equation}
1+z_{ta}= 1081\bar\delta_{int} =3.02, \:\:\:z_{ta}=2.02.
\end{equation}
We can see that the turnaround redshift based on Newtonian gravity-only evolution  is too low to match the JWST observations, but the VMOND potential lifts $z_{ta}$ to a significantly higher value.
\\\\
This is a necessary first step.

\subsection{The journey time of an outer mass shell in free fall to the mass centre}
As our next step, to produce high redshift structures such as massive galaxies and SMBH,  a turnaround overdensity needs to collapse and start virialisation as a centralised object earlier than the object's observed redshift. In \cite{bromm}, for a similar direct collapse model with Newtonian gravity, for a dark matter (including baryon) overdensity after turnaround, the dominant mode to virialisation is violent relaxation \cite{lyndenbell} with a relaxation time similar to the  free fall time to the mass centre. Our non-Newtonian model free fall time can therefore tell us whether we can expect a massive structure to appear at very high redshift.
\\\\
After $\delta\geq 1$, we can calculate the particle free fall time as follows.
The $E=h=0$ energy equation of a point around a central mass $M$ is Eq.(\ref{sdotr})
\begin{equation}
\dot{r}=Hr-\sqrt{\frac{2GM}{r}}=\sqrt{\frac{2GM}{r}}\bigg( \sqrt{\frac{H^2r^3}{2GM}}-1\bigg).
\label{dotr111}
\end{equation}
It can be argued that, when energy and angular momentum are fixed for a mass shell at turnaround $H(z) =H(z_{ta})$ is fixed during the free fall. This gives  a slightly shorter free fall time, see \cite{wong5}. Here we start with the idealised case that $H(z)$ will continue to change with time.
\\\\
In the matter dominant epoch, $H^2= 4/(9\tau^2)$. Setting
\begin{equation}
y =\frac{2}{3}\frac{r^{3/2}}{\sqrt{2GM}}, \:\:\: y=x\tau,
\label{defy}
\end{equation}
Eq.(\ref{dotr111}) simplifies to
\begin{equation}
\frac{dy}{d\tau}=\frac{y}{\tau}-1,\:\;\:\bigg(\frac{dx}{d\tau}=-\frac{1}{\tau}\bigg).
\label{yyy}
\end{equation}
Since $x$ is a dimensionless variable, one obtains a solution with a constant $\tau_0$ to be fixed.
\begin{equation}
\frac{y}{\tau}=x=\ln \bigg(\frac{\tau_0}{\tau}\bigg).
\label{eqnsoln}
\end{equation}
To fix $\tau_0$, we use the condition that, at the turnaround time $\tau=\tau_{ta}$ where $\dot{r}=0\: (\dot{y}=0)$. Then, from Eq.(\ref{yyy})
\begin{equation}
\frac{dy}{d\tau}=0, \:\:\:\frac{y(r_{ta})}{\tau_{ta}}=1=\ln\bigg(\frac{\tau_0}{\tau_{ta}}\bigg),
\end{equation}
so that $\tau_0=e\tau_{ta}$. From Eq.(\ref{eqnsoln})
as $r\rightarrow 0$  (we have  $y\rightarrow 0$) where $\tau$ goes from $\tau_{ta}$ to reach its closest approach at $\tau_{ca}=e\tau_{ta}$. Here the free fall time for a turnaround particle is
$\tau_{ff}=(e-1) \tau_{ta}=1.72\tau_{ta}$.
\\\\
The time for particles in the initial overdensity to turnaround and free fall, $\tau_{ca}$ can be given in terms of the observable redshift $z_{ca}$ as
\begin{equation}
\tau_{ca}=\tau_{ta}+\tau_{ff}=\frac{2e}{3H(z_{ta})} =\frac{2}{3H(z_{ca})},\:\:\:
\label{tvr}
\end{equation}
Using $H^2(z) =H_0^2\Omega_b(1+z)^3$ ($\Omega_b$ is the density parameter of baryon), one obtains a simple relation
\begin{equation}
 z_{ca}= 0.513(1+z_{ta})-1= 0.256 A_0-1.
\end{equation}
for $A_0$ of (\ref{dpsd}). As exemplary choices we take
\\\\
For $\delta_{int}=2.8\times 10^{-3}$, $A_0=60$, $z_{ta}=29$, $z_{ca}=15.4$. \\
For $\delta_{int}=1.8\times 10^{-3}$, $A_0=47.8$, $z_{ta}=22.9$, $z_{ca}=11.23$. \\
For $\delta_{int}=6\times 10^{-5}$, $A_0=8.47$, $z_{ta} =4.23$, $z_{ca}=1.16$.
\\\\
The first two we have already discussed. The 3rd choice corresponds to an upper limit of CMB temperature fluctuation $\Delta T/T =2\times 10^{-5}$, which is a plausible initial overdensity choice for $\sigma_8$ parameter evolution. But this overdensity reaches unity at low redshift (this result is only indicative, as the effect of dark energy will need to be taken into account).
\\\\
After $t_{ca}$, there is a central cloud with size much smaller than the overdensity cloud, where the Newtonian acceleration dominates. We can work with the Newtonian dynamical time $t_{dyn}$ to consider the phase mixing.
\\\\
This demonstrates that with overdensity greater than $2.8\times 10^{-3}$, appropriate for massive galaxies, our model could naturally produce virialised galaxies at redshift higher than $z=15$. The current highest redshift galaxy observed is MoM-z14 \cite{naidu} at $z=14.44$, which is not anticipated by $\Lambda CDM$.
\\\\
On the other hand, for low initial overdensity (e.g. $6\times 10^{-5}$), appropriate for $\sigma_{8,g}$, the $\sigma_8$ value for galaxy number density, at ($\sim 10 Mpc$) scale, calculation in our model above suggests that overdensity could reach unity at very low redshift. In \cite{carrick}$, \sigma_{8, g}=0.99$ at $z\sim0$ is observed. This is in contrast to the Newtonian perturbation theory inferred value for a Dark matter dominant overdensity which, at $z\sim 0$ is $\sigma_8=0.745$ \cite{bohringer}.  A similar baryon overdensity growth rate should also prevent baryon overdensity growing to unity, which is at odds with observation.
\\\\
This has been pursued in greater detail elsewhere \cite{wong5, wong}.

\section{Summary}
Given a central point mass in an expanding background, we have shown that it is possible to avoid the extremes of vacuole scenarios and introduce non-Newtonian mechanics in a natural way without needing to fall back onto CDM or phenomenological MOND \cite{milgrom}.  The resulting metric has both
the Schwarzschild-Lema$\hat{i}$tre metric and FLRW metric as its asymptotic solutions, with no further parameters and no additional contributions to the stress-energy tensor. This solution smoothes out  the implicit vacuoles by modifying the geodesic equation to include a non-Newtonian acceleration appropriate to the interpolating regime which we have termed VMOND.  Although the non-Newtonian potential leads to an apparent mass density in the Poisson equation, this is a purely dynamical effect and produces no pressure term. Specifically, the free fall velocity is obtained by simply adding the (negative) Newtonian free fall velocity to the velocity of the Hubble flow, reminiscent of the peculiar velocity in an expanding background in NPT, and not their accelerations. In comparison to canonical MOND \cite{milgrom} the acceleration $a_0$ which marks the boundary between Newtonian and Non-Newtonian behaviour is no longer constant, but depends on the background matter density.
\\\\
To test the formalism in its simplest form at geodesic level, an immediate consequence is that the equilibrium (or turn-around) distance $r_{TA}$
encompasses the Newtonian outer Oort cloud, which simple MOND would have as showing non-Newtonian behaviour and,
 for large galaxies and galactic clusters it relsxes the ($z$-independent) $\Lambda CDM$ bound, which is close to saturation.
\\\\
While useful first steps, the major virtue of our new baseline model is with the new solution to Einstein's Equations at galactic scales. For galactic over-density evolution at high background density and early time we find that the non-Newtonian acceleration leads to an early time over-density growth rate $\delta \propto a(t)^2$, which could lead to a high turnaround redshift $z_{ta}$ and therefore a much earlier time to start virialisation. This could provide enough time for the overdensity to virialise before $z=6.5$ which in turn, could contribute to the early appearance of massive galaxies. (JWST observations indicate that a large number of galaxies are already formed by $z=6.5$ and with no change of morphology after.) Our model provides a plausible framework to address this very high redshift massive galaxies problem. 
\\\\
It is not immediately clear that our new metric avoids anisotropy, which was Baker's \cite{baker} touchstone for a sensible metric (and which led to his choice $\eta = 3$ in contrast to our $\eta = 3/2$). In the FLRW metric, for an observer in the non-expanding comoving coordinate $\varrho$, a matter dominant universe will appear uniform and isotropic with constant density and zero pressure. Once a significant point mass at the observer is included, the underlying metric is modified. The new metric prescribes a physical distance $r=aL\neq a\varrho$ for a free falling particle shell.
As the free-fall radial distance (also for $L$) reduces, the cosmological gravitational mass inside $L$ reduces, which could lead to additional pressure. However, for an overdensity evolution at early time and late time, we notice that either in linear or Newtonian perturbation theory and in the late time pragmatic "Jeans Swindle", the evolution process is assumed adiabatic (without entropy change). This pressure term then becomes zero. We have followed this common assumption in our model. See the Appendix for further detail.
\\\\
Given the problems we have with our current cosmological models,  we could see our model as providing a new baseline for calculation.
\\\\
The title of our paper could be ironic in that, as far as we see, the metric obviates the immediate need for any conventional dark matter at galactic scales ar least, where Newtonian mechanics seems most visibly violated.
However, no single solution is complete and at universe-wide scales further gravitational sources seem required to provide the non-baryonic background potential that produces the observed CMB acoustic peaks. Within the context of this Baker model a preliminary discussion of alternative approaches to these non-baryonic sources is given elsewhere \cite{wong3}, \cite{wong2025}. In this work, we have assumed disconnection of such non-baryonic dark matter at galactic scales (i.e they do not contribute to haloes).
\\\\
Given that both $\Lambda$CDM and MOND were driven by the need to explain mass discrepancy in rotating galaxies it may seem surprising that this was not our first application of VMOND in this paper. Although it begins with the velocity modelling of equation (\ref{sdotr}) it arguably requires more details on galactic forms than we have needed here. Indeed, we can recover  the empirical Baryonic Tully-Fisher Relation \cite{wong}. This is part of a much more detailed analysis of galaxy modelling across many fronts which we have begun to perform \cite{ wong5, wong3, wong2}.

\section*{Data availability statement}
Data Sharing not applicable to this article as no datasets were generated or analysed during the current study.

\section*{Competing Interests}
The authors have no conflicts of interest to declare that are relevant to the content of this article.
\section*{Acknowledgements}
We thank Profs. Alan Heavens, Joao Magueijo, Andrew Tolley and Fran\c{c}oise Combes for helpful comments on an earlier version of the text.
\newpage

\newpage
\section{Appendix}
\subsection*{Isotropy and Non-adiabatic pressure }
As we already said, Baker's \cite{baker} insistence on isotropy as an identity in Einstein's equations forced $\eta =3$, at variance with astronomical data. We shall now argue that this was unnecessarily restrictive. In this Appendix we restrict ourselves to $\Lambda = 0$. We restore $\Lambda\neq 0$ in the main text.
\\\\
In Newtonian perturbation \cite{bosch2}, for an uniform overdensity in comoving coordinates, the pressure term due to the effect of entropy experienced by the overdensity becomes negligible if the entropy $S$ is isentropic (adiabatic). Although our parameterisation $r=a(\tau)L$ may differ from that used in the Newtonian perturbation theory, an adiabatic process would produce the same effect on the pressure term.
\\\\
To see the full implication of the adiabatic assumption, we can estimate the pressure for a mass shell at distance $r$ evolving non-adiabatically.
Since the central point mass does not associate with a pressure, the only pressure is associated with a point mass (under the influence of the central mass) free falling through the cosmological matter density.  Einstein's equation Eq.(\ref{t443}) takes a more intuitive form
\begin{equation}
-\frac{P}{c^2}=\frac{ \dot{M}(r)}{\dot{V}(r)}
\label{comovsp}
\end{equation}
where $V(r)= (4 \pi /3) r^3$, $M(r)=V(r)\rho_m$ are the volume and mass within the volume under consideration respectively. That is, when the matter flows into the volume, we have positive pressure. If matter flows out of the volume, we have negative pressure. We can answer the question we pose above and obtain the (non-adiabatic) pressure in a free falling volume due to the effective velocity $\dot{L}$. (We make the distinction since comoving volume in FLRW metric usually refers to the fixed $\varrho, \:\theta,\:\varphi$ coordinates while $a(\tau)$ is allowed to expand.) moving towards the central mass.
Within volume $V(r)$, for $\rho_m=\rho_{m,0} a^{-3}$, we have
\begin{equation}
\dot{M}=\frac{\partial}{\partial \tau} \bigg(\frac{4 \pi}{3} a^3L^3 \rho_{m,0} a^{-3} \bigg)=\frac{4 \pi}{3} 3L^3 \rho_{m,0} \frac{\dot{L}}{L}.
\end{equation}
\begin{equation}
\dot{M}=M\bigg(3\frac{\dot{L}} {L}\bigg),
\end{equation}
which states that $\dot{M}$ is effectively due to $\dot{L}/L$ in 3 directions. This can be understood by considering a point on a free falling shell moving at $\dot{r}/r$. Since a particle in the background density is moving (expanding) at the rate $H$, the effective rate that the free falling shell moves against the cosmic background density is only $\dot{L}/L=-\sqrt{2GM/r^3}$ which is due to the peculiar velocity component and describes a matter outflow at the point on the free falling shell.
\\\\
The free fall volume at $r$ remains at $V(r)$, its rate of change is given by
\begin{equation}
\dot{V}(r)=\frac{4 \pi}{3} 3r^3 \bigg(\frac{\dot{L}}{L}+ H\bigg).
\end{equation}
The effective (contracting) rate of this volume moving against the background matter density is $\dot{L}/L$, so that
\begin{equation}
-\frac{P}{c^2} =\frac{ \dot{M}(r) }{\dot{V}(r)} =\frac{M}{\frac{4\pi}{3} r^3}=\frac{V(r)\rho_m}{V(r)}=\rho_m.
\label{cosmovsp01}
\end{equation}
The anisotropy within the free falling volume is zero, since from Eq.(\ref{ni}), we have
\begin{equation}
\Delta T=T^2_2-T^1_1=\frac{r}{2} \frac{\partial T^1_1} {\partial r} =0.
\label{cosmovsp02}
\end{equation}
\\
Here the pressure $P$ is uniform and negative, there is no anisotropy.  In the $\Lambda CDM$ model where the cold dark matter is collisionless, at late time, we have $\rho_b/\rho_{\Lambda}\sim\Omega_b/\Omega_{\Lambda}=0.04/0.7$.  This pressure $P$ is due to baryons only and is about $6\%$ of the cosmological background pressure due to $\Lambda$, $P_{\Lambda} c^{-2}=-\rho_{\Lambda}$.
\\\\
In summary, without the adiabatic assumption, there is in general no anisotropy. The new-metric induced pressure due to free falling in a (real) matter dominant cosmic background tends to be small at low redshifts. However, at high redshift, this non adiabatic pressure could be significant. For example considering an overdensity cloud moving through a high matter density background at $z=100$ where radiation density is negligible, the non adiabatic pressure is now high at $Pc^{-2} \sim 10^5 P_{\Lambda}c^{-2}$. However, we can obtain this uniform negative pressure from any unspecified peculiar velocity. The implication is that either we have a generic pressure on any free falling object in a gravitationally bound system such as galaxy clusters, or this negative pressure is an artefact of our physical treatment of the averaged matter background density which may not be physically present inside the free falling volume. This is where the adiabatic assumption becomes important.
\\\\
In fact, if we can make the much stronger physical assumption of adiabatic behaviour ($\delta S/S=0$), this will ensure isotropy and no adiabatic pressure.
After an overdensity turns around $\delta \gg1$,
we note that, during galactic evolution, the central matter distribution in a galactic cluster is observed to evolve adiabatically, which leads to the postulate of the "pragmatic" Jeans Swindle \cite{jeans}-\cite{zeld}. This Jeans Swindle is used in large structure evolution simulations under MOND \cite{nusser}, \cite{sanders}. Falco et al.\cite{falco} points out that  the Jeans Swindle which states that "for an overdensity in an infinite homogeneous background, the gravitational potential is sourced by the fluctuations (overdensity) to this uniform background density, is vindicated by the right results it provides".  This means that one can again separate the background  density potential from the perturbation potential as in  Linear or Newtonian perturbation theory, and there is no pressure effect on the perturbation due to its growth (and contraction) in the matter density background. We can argue that at late time a free falling particle does not physically encounter any mean matter density and does not generate entropy and therefore the adiabatic approximation can still hold. However, this phenomenological Jeans Swindle has no formal justification.
\\\\
We shall assume that the adiabatic approximation at early time continues to hold at late time and there is no pressure term experienced by a free falling particle. Isotropy is preserved.

\begin{thebibliography}{99}
%
%
\bibitem{milgrom} Milgrom, M., The Astrophysical Journal, 270, 371-383 (1983). https://doi.org/10.1086/161130
%
\bibitem{mcgaugh2024} McGaugh, S. S., Schombert, J. M., Lelli, F.,  Franck, J., Accelerated Structure Formation: the Early Emergence of Massive Galaxies and Clusters of Galaxies. 	Accepted for publication in the Astrophysical Journal.
https://doi.org/10.48550/arXiv.2406.17930
%
\bibitem{baker} Baker, G. A. Jr., Effects on the structure of the universe of an accelerating expansion. General Relativity and Gravitation. 34 (6), 767-791 (2002)  https://doi.org/10.1023/A:1016371629024
%
\bibitem{mcgaugh2004} McGaugh, S. S., The Mass Discrepancy-Acceleration Relation: Disk Mass and the Dark Matter Distribution. APJ, 609, 652-666, (2004); https://doi.org/10.1086/421338
\\
McGaugh, S. S., The Baryonic Tully-Fisher Relation of Galaxies with Extended Rotation Curves and the Stellar Mass of Rotating Galaxies. APJ. 632, 859-871 (2005).  https://doi.org/10.1086/432968
\\
McGaugh, S. S., The Baryonuc Tully-Fisher Relation of gas-rich Galaxies as a test of $\Lambda$CDM and MOND. AJ, 143:40 (15pp.) (2012) https://doi.org/10.1088/0004-6256/143/2/40
%
\bibitem{lelli} Lelli, F., McGaugh, S. S., Schombert, J.M., Pawlowski, M.S.,
One Law to Rule Them All: The Radial Acceleration Relation of Galaxies.
 Astrophysical Journal. 836, 152 (2017). https://doi.org/10.3847/1538-4357/836/2/152
 %
 \bibitem{weinberg} Weinberg DH, Bullock JS, Governato F, Kuzio de Naray R, Peter AH. Cold dark matter: Controversies on small scales. Proc Natl Acad Sci U S A. 2015 Oct 6;112(40):12249-55. doi: 10.1073/pnas.1308716112. Epub 2015 Feb 2. PMID: 25646464; PMCID: PMC4603506.
%
\bibitem{riess} L. Verde, T. Treu,  A. Riess, "Tensions between the
Early and the Late Universe", arXiv:1907.10625.
%
\bibitem{valentino}  E. Di Valentino et al., "Cosmology intertwined II: The Hubble
constant tension", arXiv:2008.11284.
%
\bibitem{bohringer} Bohringer, H., Chon, G., Collins, C. A., 2014, Astron. Astrophys. 570, A31.
https://doi.org/10.3847/1538-4357/ab12d6
%
\bibitem{einasto} J. Einasto et al., "Steps toward the Power Spectrum of Matter. II. The Biasing Correction with $\sigma_8$ Normalization", Astrophysical Journal, 519, 456-468 (1999).
https://doi.org/10.1086/307385
%
\bibitem{karim} M. Abdul Karim et al., "DESI DR2 Results II: Measurements of Baryon Acoustic Oscillations and Cosmological Constraints", 	Phys. Rev. D 112, 083515, 2025,
https://doi.org/10.1103/tr6y-kpc6.
%
\bibitem{ozulker} E. $\ddot{O}$z$\ddot{u}$lker, E.  Di Valentino, W. Giar$\acute{e}$,  "Dark Energy Crosses the Line: Quantifying and Testing the Evidence for Phantom Crossing",
 https://doi.org/10.48550/arXiv.2506.19053
%
\bibitem{ferreira} Ferreira, L., et al., Panic! At the Disks: First Rest-frame Optical Observations of Galaxy Structure at $z > 3$ with JWST in the SMACS 0723 Field, Astrophysical Journal Letters 938 L2 https://doi.org/10.3847/2041-8213/ac947c
%
\bibitem{yan} Yan, H., et al., First Batch of Candidate Galaxies at Redshifts 11 to 20 Revealed by the James Webb Space Telescope Early Release Observations,
https://doi.org/10.3847/2041-8213/aca80c
%
\bibitem{ferreira2} Ferreira, L., The JWST Hubble Sequence: The Rest frame Optical Evolution of Galaxy Structure at $1.5<z<6.5$, Astrophysical  Journal  955, 2 (2023). https://doi.org/10.3847/1538-4357/ace76
%
\bibitem{stefano} Stefano, C., et al.   A shining cosmic dawn: spectroscopic confirmation of two luminous galaxies at $z\approx 14$, https://doi.org/10.48550/arXiv.2405.18485
%
\bibitem{naidu} Naidu, R. P., et al., A Cosmic Miracle: A Remarkably Luminous Galaxy at $z_{spec}=14.44$ confirmed with JWST, Submitted to Open Journal of Astrophysics,
https://doi.org/10.48550/arXiv.2505.11263
%
\bibitem{bertone} Bertone, G., Tait, T.M.P. A new era in the search for dark matter. Nature 562, 51-56 (2018). https://doi.org/10.1038/s41586-018-0542-z
%
\bibitem{nusser} Nusser, A., Modified Newtonian dynamics of large-scale structure, MNRAS, 331, 909-916 (2002). https://doi/org/10.1046/j.1365-8711.2002.05235.x
%
\bibitem{boran} Sibel Boran, S., Desai, S., Kahya, E.,Woodard, R., GW170817 Falsifies Dark Matter Emulators, Phys. Rev. D 97, 041501 (2018), https://doi.org/10.1103/PhysRevD.97.041501
%
\bibitem{moffat} Moffat, J.W., Acceleration in Modified Gravity (MOG) and the Mass-Discrepancy Baryonic Relation,  arXiv:1610.06909v2. https://doi.org/10.48550/arXiv.1610.06909
%
\bibitem {skordis} Skordis, C., Zlosnik, T., Gravitational alternatives to dark matter with tensor mode speed equaling the speed of light. Phys. Rev. D 100, 104013, https://doi.org/10.1103/PhysRevD.100.104013
%
\bibitem{skordis1} Skordis, C., Zlosnik, T.,  A new relativistic theory for Modified Newtonian Dynamics, Phys. Rev. Lett. 127, 161302 (2021), https://doi.org/10.1103/PhysRevLett.127.161302
%
\bibitem{cao} Cao, Z., et al. (The LHAASO Collaboration),  Stringent Tests of Lorentz Invariance Violation from LHAASO Observations of GRB 221009A, Phys. Rev. Lett. 133, 071501 (2024), https://doi.org/10.1103/PhysRevLett.133.071501
    %
\bibitem{banik} Banik, I., et al., Strong constraints on the gravitational law from Gaia DR3 wide binaries.
https://doi.org/10.1093/mnras/stad3393
%
\bibitem{tremaine} Vokrouhlický, D., Nesvorný, D., Tremaine, S., Testing MOND on small bodies in the remote solar system, https://doi.org/10.48550/arXiv.2403.09555
%
\bibitem{einstein} Einstein, A., Straus, E., The influence of the Expansion of Space on the gravitational Fields surrounding the individual stars, Rev. Mod. Phys. 17, 120-124 (1945). https://doi.org/10.1103/RevModPhys.17.120
%
\bibitem{stuchlik} Stuchlik, Z., An Einstein-Strauss-De-Sitter Model of the Universe, Bull. Astron. Inst. Czechosl. 35, 205-215 (1984).
Publisher, Academia, Praha,
Publication country,
Czechoslovakia.
%
\bibitem{balbinot} Balbinot, R., Bergamini, R., Comastri, A., Solution of the Einstein-Strauss problem with a $\Lambda$ term Phys. Rev. D38,  2425 (1988). https://doi.org/10.1103/PhysRevD.38.2415
%
\bibitem{carrera} Carrera, M., Giulini, D., On the influence of the global cosmological expansion on the local dynamics in the Solar System, Rev. Mod. Phys. 82, 169-208 https://doi.org/10.1103/RevModPhys.82.169
%
\bibitem{mcvittie} McVittie, G. C., The Mass-Particle in an Expanding Universe. MNRAS, 93, 325-339 (1933). https://doi.org/10.1093/mnras/93.5.325
%
\bibitem{nolan} Nolan, B. C., A point mass in an isotropic universe. Existence, uniqueness and basic properties, Phys. Rev. D58, 064006 (1998), https://doi.org/10.1103/PhysRevD.58.064006
%
\bibitem{pavlidou} Pavlidou, V., Tomaras,  T. N., Where the world stand still: turnaround as a strong test of $\Lambda$CDM cosmology,
https://doi.org/10.1088/1475-7516/2014/09/020
%
\bibitem{bona} Bona, C., Stela,  J., 'Swiss cheese' models with pressure. Phys. Rev. D36, 2915-2918 (1987). https://doi.org/10.1103/PhysRevD.36.2915
%
\bibitem{dehaas} de Haas, P., Constructing Spacetime from Rapidity: Explicit BQ Derivation of the PG coframe and the metric tensor, DOI:10.5281/zenodo.17717365
%
\bibitem{mukhanov} Mukhanov, V., Physical foundation of cosmology, Cambridge University Press 2005.
%
\bibitem{krasinski} Bolejko, K.,  C$\acute{e}$l$\acute{e}$rier M-N., Krasi$\acute{n}$ski ,A., Inhomogeneous cosmological models: exact solutions and their applications, Class. Quantum Grav. 28, 164002 (2011).
%
\bibitem{khoury} Berezhiani, L.,  Khoury, J., Theory of dark matter superfluidity, Phys. Rev. D92,  103510 (2015)
%
\bibitem{bosch2} Van den Bosch, F., Theory of Galaxy Formation, Lecture 4, Yale University (Fall 2022).
%
%
\bibitem{mitra} Mitra, A., Friedmann-Robertson-Walker metric in curvature coordinates and its applications, Grav. cosmol. No.2 (2013).
%
\bibitem{mitra2} Mitra, A., Interpretational conflicts between static and non-static forms if the de-Sitter metric, Nature, Scientific reports 2, article number:923 (2012).
%
\bibitem{binney} Binney,  J., Tremaine, S., Galactic Dynamics, Princeton University Press, (2008).
%
\bibitem{bromm} Bromm, V., Loeb,  A., ApJ, 596, 34 (2003),
https://doi.org/10.1086/377529
%
\bibitem{hwang} Hwang, J., Noh., H., Newtonian, post-Newtonian and Relativistic Cosmological
Perturbation Theory, Nuclear Physics B Proceedings Supplement 00 (2018) 1–7, arxiv.org/pdf/1509.04703.
%
\bibitem{fan} Fan, X.,  NewA Rev., 50, 665 (2006)
%
\bibitem{miyatake} Miyatake, H., et al., First Identification of a CMB Lensing Signal Produced by 1.5 Million Galaxies at $z\approx 4$: Constraints on Matter Density Fluctuations at High Redshift,
https://doi.org/10.1103/PhysRevLett.129.061301
%
\bibitem{milgrom1986} Milgrom, M., ApJ, 306, 9 (1986). DOI: 10.1086/164314
%
\bibitem{oria} Oria, P., et al., The phantom dark matter halos of the Local Volume in the context of modified Newtonian dynamics,
https://doi.org/10.3847/1538-4357/ac273d
%
\bibitem{sanders} Sanders, R. H., Forming galaxies with MOND. MNRAS, 386, 1588-1596 (2008). https://doi.org/10.1111/j.1365-2966.2008.13140.x
%
\bibitem{barkana} Barkana, R., Loeb, A., In the Beginning: The First Sources of Light and the Reionization of the
    Universe, Phys. Rept. 349, 125-238, 2001.1310.1920
https://doi.org/10.1016/S0370-1573
%
\bibitem{lyndenbell} Lynden-Bell, D., Statistical Mechanics of Violent Relaxation in Stellar Systems, MNRAS, 136, 101-121 (1967). DOI: 10.1093/mnras/136.1.101
%
\bibitem{wong5} Wong, C. C., Variable Modified Newtonian Mechanics II: Non Rotating Galaxies,
DOI:10.13140/RG.2.2.21479.16808
%
\bibitem{carrick} Carrick, J. et al., "Cosmological parameters from the comparison of peculiar velocities with predictions from the 2M++ density field", arXiv:1504.04627, https://doi.org/10.1093/mnras/stv547.
%
\bibitem{wong} Wong, C. C., Variable Modified Newtonian mechanics III: Milky Way Rotational Curve, arXiv:1802.01493, https://doi.org/10.48550/arXiv.1802.01493
%
\bibitem{wong3} Wong, C. C., Variable Modified Newtonian Mechanics V: Cosmic Microwave Background Angular Power specttrum, DOI:10.13140/RG.2.2.13740.95368
%
\bibitem{wong2025} Wong, C. C., Variable Modified Newtonian Mechanics VII: Supermassive Black Hole at very high redshift,
https://doi.org/10.48550/arXiv.2506.05423
%
\bibitem{wong2} Wong, C. C., Variable Modified Newtonian Mechanics IV: Turnaround radius, arXiv:1910.10477
https://doi.org/10.48550/arXiv.1910.10477
%
\bibitem{jeans} Jeans, J. H., The Universe Around Us. Cambridge University Press (1929).
%
\bibitem{zeld} Zeldovich, Y. B., Novikov, I.D., Relativistic astrophysics
Vol.1: Stars and relativity, University of Chicago Press, 1971.
%
\bibitem{falco} Falco, M., et al. Why does the Jeans Swindle work?. MNRAS 2013, 431, pp.L6-L9. https://doi.org/10.1093/mnrasl/sls051



\end{thebibliography}
\end{document}